\documentclass[epj]{svjour}
\usepackage{graphicx}
\usepackage{amsmath}
\usepackage{multirow}
\usepackage{color}

\newcommand{\D}[0]{\ \mathrm{d}}

\newcommand{\Ca}[0]{\mathrm{Ca}}
\newcommand{\V}[1]{\boldsymbol{#1}}
\begin{document}
\title{Oblate to prolate transition of a vesicle under flow}
\author{Maximilien Degonville\inst{1} \and Gwenn Boedec\inst{1} \and Marc Leonetti\inst{2}
}                     
\offprints{gwenn.boedec@univ-amu.fr, marc.leonetti@univ-grenoble-alpes.fr}          
\institute{Aix Marseille Univ, CNRS, Centrale Marseille, IRPHE UMR 7342, 13384, Marseille, France \and Univ. Grenoble Alpes, CNRS, Grenoble INP, LRP, Grenoble, France}
\date{Received: date / Revised version: date}
%
\abstract{
Vesicles are micrometric soft particles whose the membrane is a two-dimensional incompressible fluid governed by bending resistance leading to a zoology of shapes. The dynamics of deflated vesicles in shear flow with a bottom wall, a first minimal configuration to consider confined vesicles is investigated using numerical simulations. Coexistence under flow of oblate (metastable) and prolate (stable) shapes is studied in details. In particular, we discuss the boundaries of the region of coexistence in the ($v,Ca$) plane where $v$ is the reduced volume of the vesicle and $Ca$ the Capillary number. We characterize the transition from oblate to prolate and analyse the divergence of the transition time near the critical capillary number. We then analyse lift dynamics of oblate vesicle in the weak flow regime.
} 
\maketitle
\section{Introduction}
\label{intro}

Dynamics of vesicles under flow has triggered numerous studies in the past decades, motivated in part by the connection with red blood cells dynamics in blood flow since their interfacial mechanics share similar properties.
Indeed, both are phospholipidic bags, separating an inner viscous fluid from the suspending outer fluid.
Vesicles are also a member of a broad family of soft objects which may be described loosely as droplets with ``complex'' interfacial properties, starting from clean drops described by a constant surface tension and extending to capsules (polymerized elastic interface) \cite{BarthesBiesel_2016} or polymersomes (vesicles for which the membrane shear viscosity plays a dominant role) \cite{Dimova_2002,Discher_2006,Meng_2009}.
Studies are generally interested in how the coupling between hydrodynamics and interfacial mechanics selects the dynamics under flow: for instance, when submitted to a simple shear flow, both vesicles \cite{Kantsler_2005} and capsules \cite{DeLoubens_2016} exhibit a tank-treading motion consisting of a fixed ellipsoidal shape and a rotation of the interface along this fixed shape, as well as other type of motions, such as tumbling and trembling when internal viscosity is increased \cite{Deschamps_2009}.

Another question of particular interest is how the deformability of these objects can lead to cross-streamline migration, in particular in the context of a wall-bounded flow, a simple configuration which allows to determine the leading-order contribution of channel's wall-induced migration.
Indeed, for these objects, fluid flows are generally in the creeping regime. In this regime, the equations governing fluid flows are linear and reversible.
Thus, a symmetric rigid body submitted to a shear flow near a wall does not migrate away from the wall.
On the contrary, if the body is deformable, the symmetry with respect to the flow may be broken and will induce a lift velocity pushing the particle away from the wall, as first observed experimentally for drops in \cite{Goldsmith_1962}.

Far from the wall, the migration velocity can be related to the particle stresslet and has been shown theoretically to scale as $v_{lift} \sim \dot{\gamma} R_0^3 / h^2$ \cite{Smart_1991}. This far-field scaling can be recovered either by approximating the shape of the vesicle by an ellipsoid \cite{Olla_1997}, or by expansion around the spherical shape \cite{Vlahovska_2007,Farutin_2013}. This far-field scaling was confirmed numerically \cite{Zhao_Pof2011,Farutin_2013} and measured experimentally under microgravity \cite{Callens_2008}.
Close to the wall, the question of how the hydrodynamic lift force can trigger unbinding of an adhering  vesicle was tackled numerically and analytically in two \cite{Cantat_1999} and three \cite{Seifert_1999,Sukumaran_2001} dimensions, and also investigated experimentally \cite{Lorz_2000,Abkarian_2002,Abkarian_2005}. Gravity effects were also investigated numerically \cite{Sukumaran_2001,Messlinger_2009,Zhao_Pof2011}. While a deflated vesicle close to a wall is expected to have a flattened shape due to gravity or adhesive interaction \cite{Kraus_1995}, most of the focus has been devoted to the dynamics of elongated shapes.

Indeed, among soft objects, a notable peculiarity of vesicles is that multiple (locally stable) shapes at thermal equilibrium may exist, among them the prolate (cigar-like) and the oblate (disk-like) shapes \cite{Seifert_1991,Seifert_1997}. Which family is the globally stable equilibrium shape depends only on the reduced volume : for quasi-spherical shapes, prolate shapes are the one with the least bending energy, while for more deflated vesicles, oblate shapes are the global minimum. In this respect they offer a unique system to study the influence of the initial condition on the dynamics and how external forcing by hydrodynamics may lead to a transition from one shape to the other. Most of the aforementioned works were devoted to the study of prolate vesicles, especially in the near-spherical regime $v \in [0.9,1]$. In the context of wall-bounded shear flow, Zhao et al. \cite{Zhao_Pof2011} considered oblate shapes for a reduced volume of $v=0.65$ for one value of shear rate, but the transition to prolate was not studied. In the context of free space shear flow, Spann et al. \cite{Spann_2014} studied the phase diagram of an initially biconcave vesicle as a function of dimensionless shear rate and viscosity contrast $\lambda$ for two values of reduced volume. They found that for a given capillary number, increasing the viscosity contrast $\lambda$ stabilizes the biconcave shape, which otherwise may transition towards prolate shape. For a reduced volume of $0.65$, and at low capillary numbers, biconcave shape was metastable for all viscosity contrast. Increasing the capillary number while keeping viscosity contrast fixed leads to a transition to prolate shapes, occurring for $\lambda=1$ at a capillary number in the range 0.6-1.0. On the other hand, for a reduced volume of $0.75$, a viscosity contrast higher than a value in the range 3.5-4 was necessary to stabilize the biconcave shape.
Noguchi and Gompper \cite{Noguchi_2004,Noguchi_2005} studied a similar phase diagram for a reduced volume of $0.59$ but with membrane viscosity instead of viscosity contrast, and found that including membrane viscosity also stabilizes the oblate shape. Note that none of the aforementionned studies characterized the nature of the transition, nor studied its phase diagram in $(v,Ca)$ space.

In this paper, we investigate how the coexistence of oblate and prolate shapes is altered by the hydrodynamic stresses, and characterize the nature of the transition. We study how the critical shear rate depends on the reduced volume, and provide a phase diagram in the $(v,Ca)$ space. As an example where this bistability has physical consequences, we study lift dynamics close to wall. The paper is organized as follows: first, we describe the model used and its transcription into a numerical model. Then, we focus on the hydrodynamic-induced transition from oblate to prolate shapes.
We then analyse lift dynamics of oblate vesicles, both below and above critical shear rate. We conclude with some observations of high capillary number lift dynamics.

\section{Model and numerical method}
\begin{figure}
\includegraphics[scale=0.85]{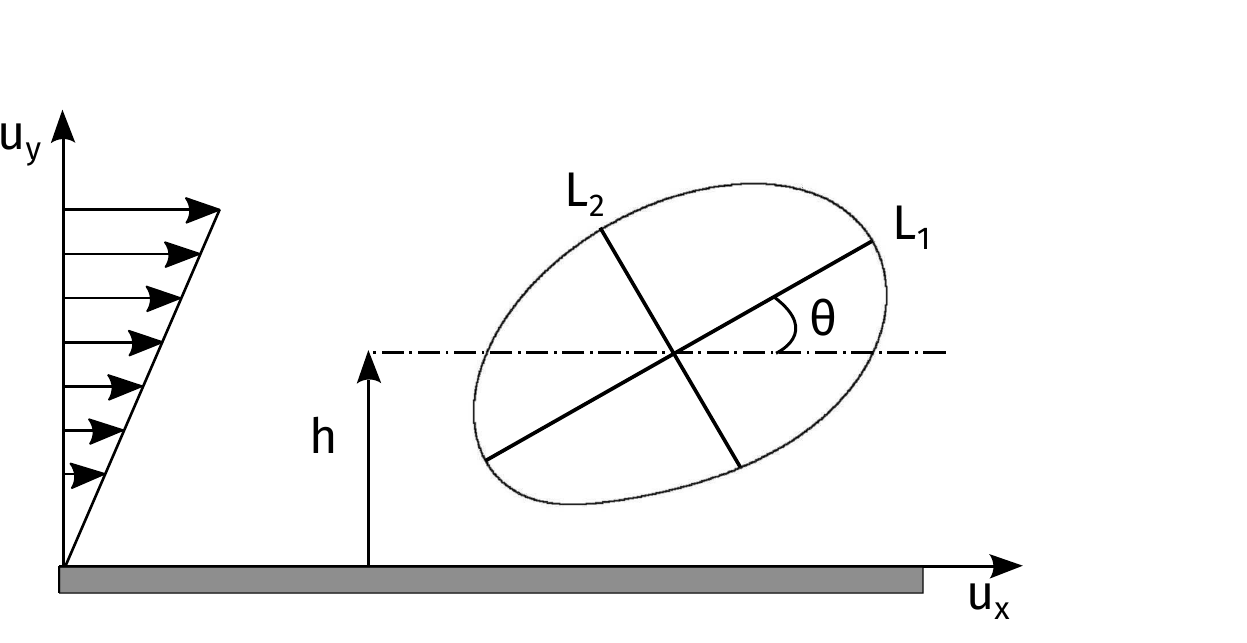}
\caption{Sketch of the situation considered here : an ellipsoidal vesicle is placed in a wall bounded shear flow. The deformation is characterized by the Taylor parameter $D$: here, with our choice of axis, $D_{xy}=(L_1-L_2)/(L_1+L_2)$.  }
\label{fig_scheme}
\end{figure}
A vesicle is a drop enclosed by a self-organized fluid phospholipid membrane which possesses peculiar properties : lacking a reference configuration, the membrane behaves as a bi-dimensionnal incompressible fluid on a curved surface. There is however an energetic cost associated with the curvature of the membrane, measured by the Helfrich free energy :
\begin{equation}
E^{\kappa}  = \frac{\kappa}{2}\int_S(2H)^2 \D S 
\end{equation}
where $\kappa \approx  20 \ k_B T$ is the bending modulus and $H$ is the mean curvature of the surface $S$. The dynamics of the vesicle under flow thus results from the interplay between the elastic forces $\V{f^{\kappa}}$ related to bending energy such as $\delta E^\kappa = - \int_S \V{f^{\kappa}} \cdot \delta \V{x} \D S$, and the fluid stresses related to the surrounding (inner and outer) media. An additional specificity of this phospholipidic membrane is that surface flows are locally incompressible to preserve the density of phospholipids. This induces an additional stress in the form of a bidimensionnal pressure (homogeneous to a surface tension) $\gamma$, which is a Lagrange multiplier of the surface divergence free constraint
\begin{equation}
\V{\nabla_S} \cdot \V{v}^{m} = 0
\end{equation}
where $\V{\nabla_S}$ is the surface divergence, and $\V{v}^m$ is the membrane velocity field. This membrane velocity field is the restriction on the membrane of the surrounding velocity fields $\V{v}^m(\V{x})=\V{v}^i(\V{x})=\V{v^e}(\V{x})$, which are governed by the Stokes equation
\begin{equation}
\eta^{i,e} \Delta \V{v}^{i,e} - \V{\nabla} p^{i,e} = \V{0} \ , \ \V{\nabla}\cdot\V{v}^{i,e}= 0
\end{equation}
where $\eta^{i,e}$ are the internal and external viscosities.\\
Far from the wall, the velocity is imposed as a simple shear flow
\begin{equation}
\V{v^e}(\V{x}) \rightarrow_{|\V{x}| \rightarrow \infty} \V{v}^{\infty} = \dot{\gamma} y \V{e_x}
\end{equation}
Examples of faccurate codes for three-dimensional simulations of vesicles in flow were published in 2011 \cite{Boedec2011,Zhao2011,Biben2011}. Here the surface meshing is different. To solve this set of equations, we introduce a discretization of the interface $S$ as a collection of Loop subdivision elements. With this discretization, bending membrane forces can be computed using a finite element method. This loop elements have also been used in the case of capsules and droplets with surface viscosities \cite{Loubens2015,Gounley2016}. The Stokes equations for inner and outer fluid are solved by a boundary element method which, in the case of equal inner and outer viscosities, relates the velocity on the interface to the force densities via an integral relation :
\begin{equation}
v_i(\V{x}) = v_i^{\infty}(\V{x}) + \frac{1}{8\pi\eta} \int_S G^w_{ij}(\V{x},\V{x'}) f^m_j (\V{x'}) \D S (\V{x'})
\label{eq_bem}
\end{equation}
where $G^w_{ij}$ are the cartesian components of the Green function taking into account the presence of the wall at $y=0$ \cite{Blake_1971}, and $f^m_j$ are the cartesian components of the membrane forces $\V{f}^m = \V{f}^\kappa+\V{f}^{\gamma}$. This equation is computed at each node of the mesh describing the interface. The Lagrange multiplier $\gamma$ is determined by imposing that $\V{\nabla_s} \cdot \V{v} = 0$ at each node of the mesh. Finally, the interface position is advanced by using a trapezoidal method 
\begin{equation}
\V{x}^{t+dt} = \V{x}^t + \frac{dt}{2}\left[\V{v}(\V{x}^t)+\V{v}(\V{x}^{t+dt})\right]
\end{equation}
where the unknown position $\V{x}^{t+dt}$ is found by an iterative method.
More details and validation of the numerical method may be found in \cite{Boedec_2017}.

The dynamics of a vesicle under shear flow is described by two dimensionless numbers, the reduced volume $v$ and the capillary number $\Ca$ defined by
\begin{equation}
v\,=\,\frac{V}{\frac{4}{3}\pi(\frac{S}{4\pi})^{3/2}}
\end{equation}
where $S,V$ are the area and the volume of the vesicle.
\begin{equation}
Ca\,=\,\frac{\eta\dot{\gamma}R_0^3}{\kappa}
\end{equation}
where $R_0 = \left(\frac{3V}{4\pi}\right)^{1/3}$ is the radius of the sphere having the same volume.\\
To study the influence of the initial shape on the wall migration, we set deflated vesicles at a distance $h=1.1 R_0$ of the wall. The initial shape is obtained by letting an oblate/prolate ellipsoid relax ($\Ca=0$) according to (\ref{eq_bem}). We consider reduced volumes in the interval $[0.59,0.75]$. Since both oblate and prolate equilibrium shapes exist for reduced volumes in this interval, starting with an oblate (resp. prolate) ellipsoid, the shape relaxes towards the oblate (resp. prolate) thermodynamic equilibrium shape. The shear flow is then started abruptly, and the evolution of the vesicle is monitored (see figure \ref{fig_scheme}): we measure $h$, the vertical position of the center of gravity of the vesicle, the axis $L_1,L_2,L_3$ of the ellipsoid having the same tensor of inertia, and the angle of inclination $\theta$ of the shape. If prolate shape is the minimum global, an initial condition of oblate shape can still be obtained experimentally before starting flow if the vesicle settles to the bottom of the chamber with a well chosen Bond number: see the diagram, figure 1 of Kraus et al \cite{Kraus_1995}.


\section{Oblate to prolate transition}

\begin{figure}
\includegraphics[scale=0.44,trim=5 210 0 220, clip=true]{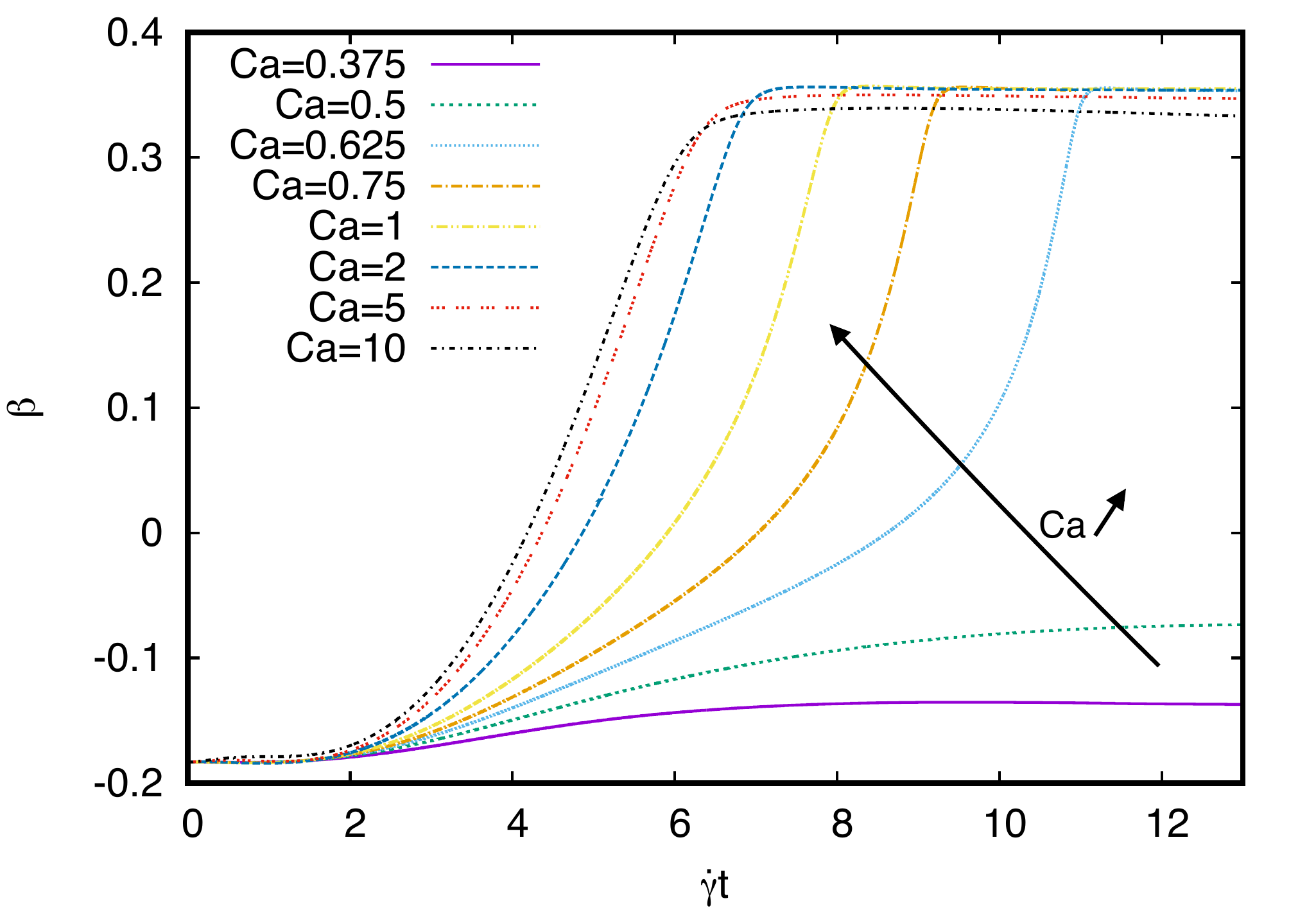}
\caption{Evolution of the shape parameter as a function of dimensionless time, for different capillary numbers, and for a $v=0.635$ vesicle.}
\label{fig_beta_time}
\end{figure}

Before looking into the influence of the initial shape on the dynamics, we characterize the zone for which two possible long term solutions can exist, and the nature of the transition between these two states. To do so, we study the evolution of an initially oblate vesicle submitted to shear flow for various capillary numbers. As will be shown at the end of this section, the presence of the wall affects only slightly the quantitative results for the transition, and thus the results presented in this section is applicable in both cases. 

\begin{figure*}
\center
\includegraphics[scale=0.65,trim=0 335 25 0, clip=true]{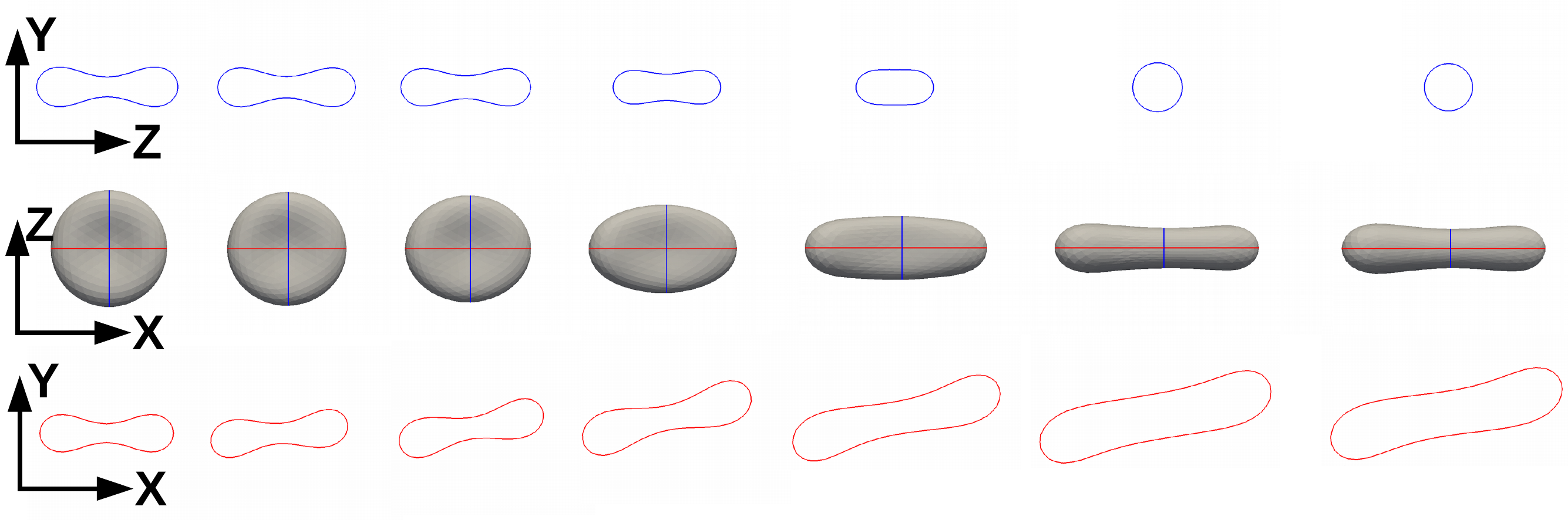}
 \caption{Oblate to prolate transition triggered by hydrodynamic stresses ($v=0.635, Ca=2$). Snapshots in the $Z-Y$ plane (upper row), in the $X-Z$ plane (middle row) and in the $X-Y$ plane (bottom row) for different dimensionless times (from left to right $\dot{\gamma} t = 0;1;2;4;6;8;12;$). } 
 \label{fig_transition_oblate_prolate}
\end{figure*}
\begin{figure*}
\center
\begin{tabular}{cc}
\includegraphics[scale=0.7]{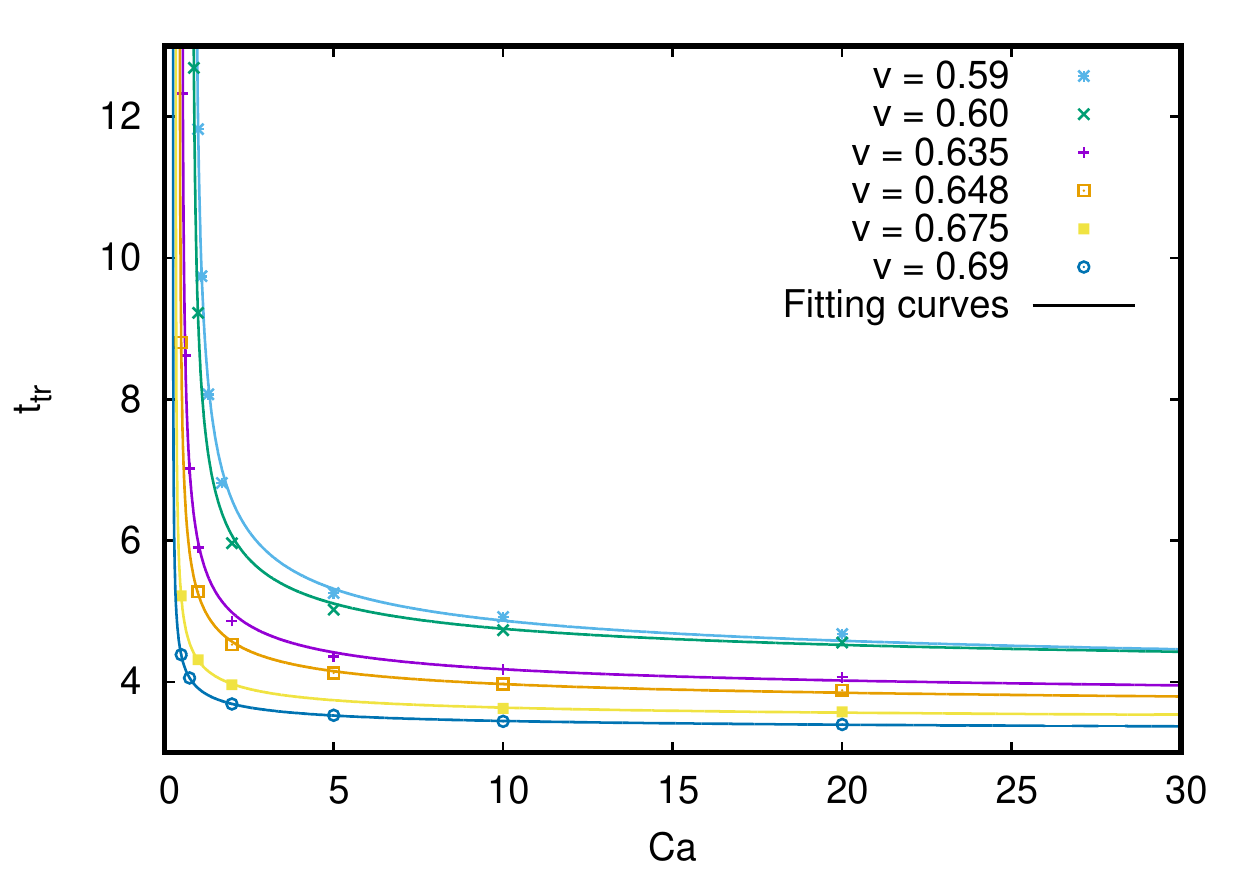}&
\includegraphics[scale=0.7]{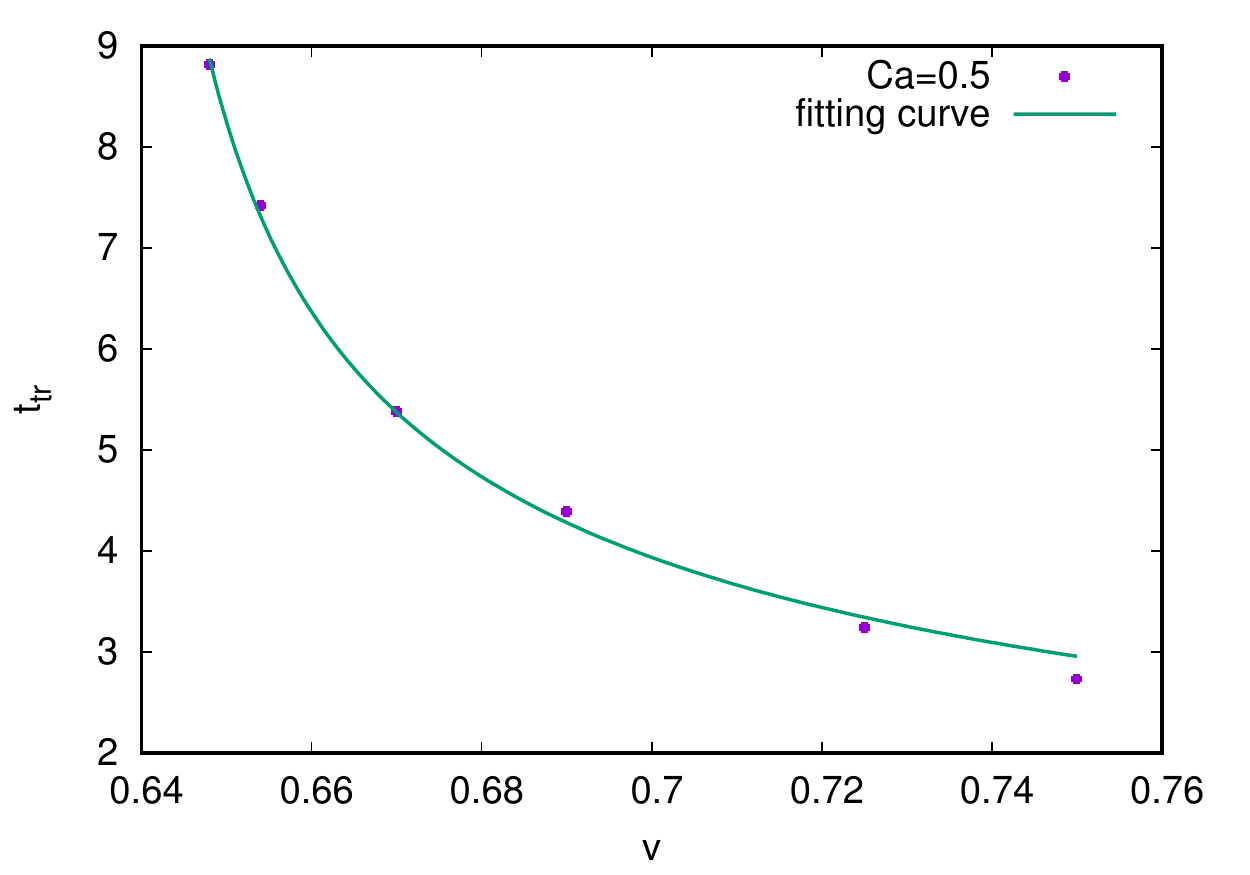}\\
(a) & (b)\\
\end{tabular}
\caption{Evolution of the transition time as a function of capillary number, for different reduced volume. Symbols correspond to simulation points, solid lines corresponds to fit by $1/\sqrt{Ca-Ca_c}+t_{\infty}$. This behavior is characteristic of a saddle-node bifurcation. $t_{\infty}$ corresponds to the time needed to reorient the vesicle in the flow. (b) Evolution of the transition time as a function of reduced volume for a fixed capillary number ($Ca=0.5$). Symbols correspond to simulation points, solid lines corresponds to fit by $1/\sqrt{v-v_c}$. }
\label{fig_transitionTime_Ca}
\end{figure*}

We introduce a shape parameter $\beta = (L_1/R_0-1)(L_2/R_0-1)(L_3/R_0-1)$ where the $L_i$ are the  axis of the ellipsoid having the same tensor of inertia as the vesicle. Furthermore, $L_1L_2L_3=R_0^3$ due to the constraint on the volume. Thus, the shape parameter allows to distinguish between prolate $(\beta>0)$ and oblate $(\beta<0)$ shapes. Figure \ref{fig_beta_time} shows the evolution of this shape parameter as a function of dimensionless time $\dot{\gamma}t$, for different capillary numbers. Two regimes clearly emerge: for weak flows $(Ca<Ca_c)$, an initially oblate shape is slightly deformed by hydrodynamic stresses, but remains oblate $(\beta<0)$ for all times. By contrast, under sufficient hydrodynamic stresses $(Ca>Ca_c)$, an initially oblate shape will be stretched by the flow and will eventually transition towards a prolate shape $(\beta>0)$. This transition is illustrated in figure \ref{fig_transition_oblate_prolate}, where a side and a top view of the vesicle are provided. After the flow is started, the shape is stretched in the shear plane. Because volume and surface are invariants of the dynamics, this stretching is accompanied by a thinning of the shape in the $Z$ direction, perpendicular to the flow. The dynamics of the profile in the $Z-Y$ cross section is interesting to understand the transition: just before the transition, the cross profile which had a characteristic ''discocyte`` shape, passes through a ''pancake`` shape which is convex (while the discoctye was not). Looking solely at bending forces in this plane, it is clear that a convex shape such as the one shown at $t=6$ in figure \ref{fig_transition_oblate_prolate} is not stable and will evolve towards a circular shape. This is because, in this plane which is perpendicular to the flow plane, the flow acts only indirectly, through the constraint of volume and surface preservation. Thus, there is no external flow that can counter the effect of bending forces in this plane, which leads to the evolution towards circular profile.

\begin{figure*}
\begin{tabular}{cc}
\includegraphics[scale=0.71,trim = 0 10 0 0,clip=true]{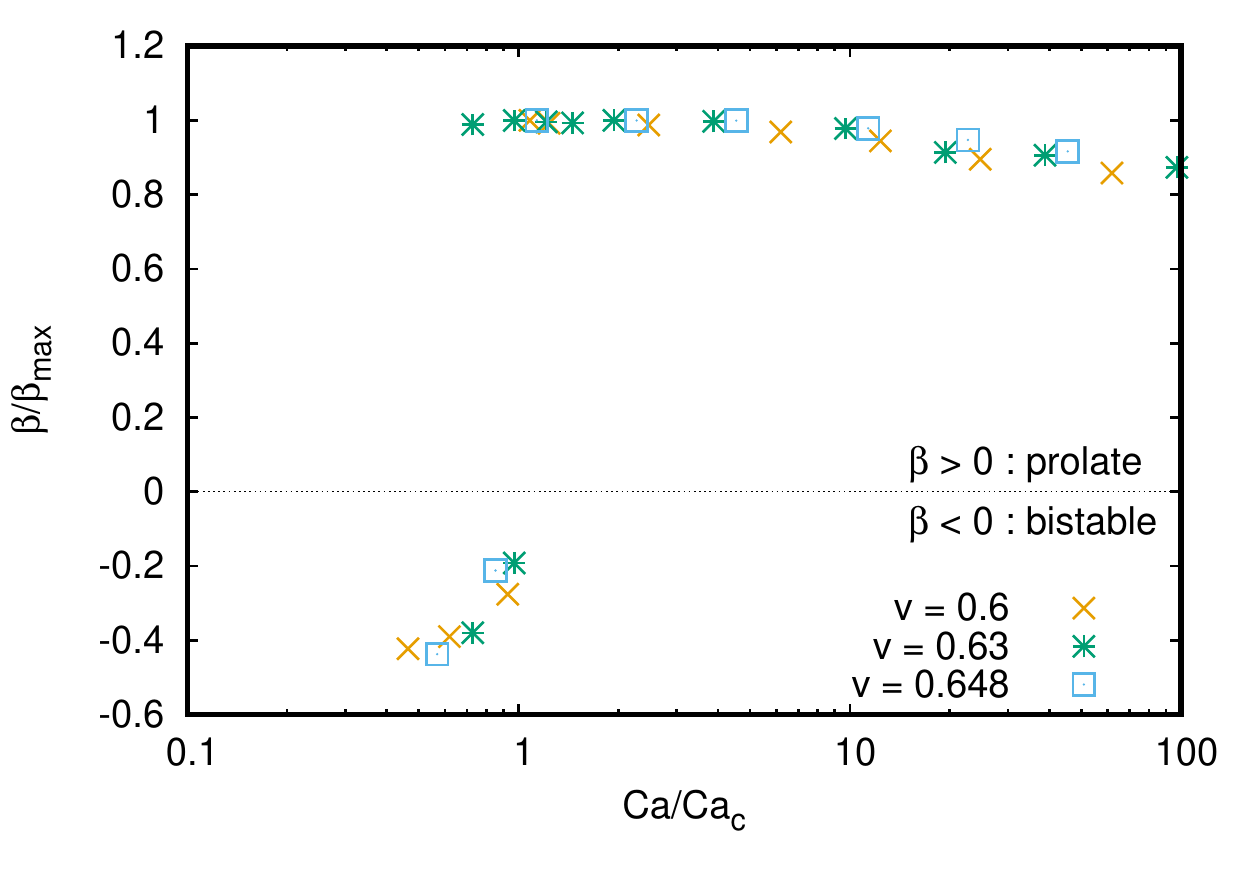}&
\includegraphics[scale=0.68]{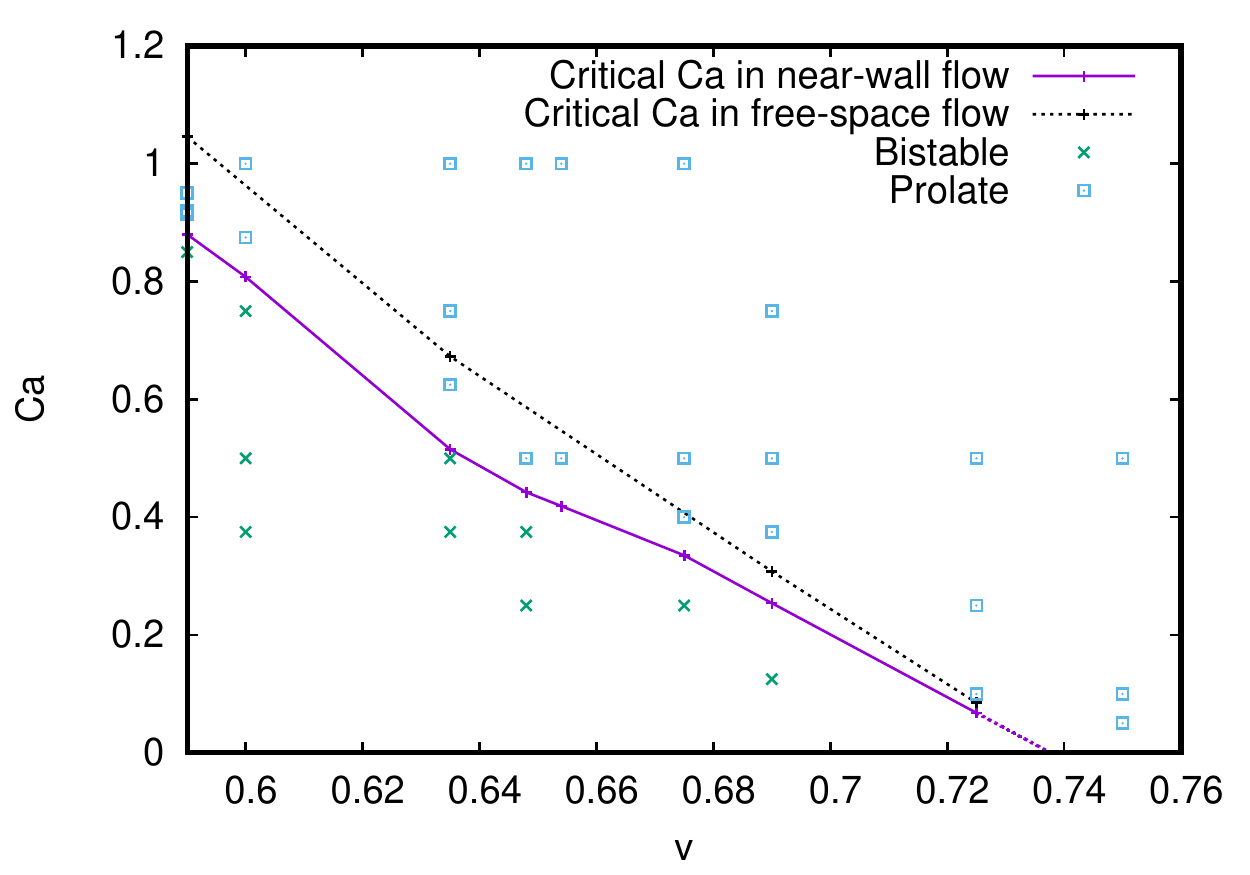}\\
(a) & (b)\\
\end{tabular}
\caption{(a) Evolution of the shape parameter (rescaled by maximal value) as a function of capillary number rescaled by critical capillary number, for different reduced volume. All points collapse to two branches of solutions, one prolate ($\beta/\beta_{max}\approx 1$), one oblate ($\beta/\beta_{max}<0$).  (b) Critical capillary number separating the bistable region from the prolate region as a function of reduced volume. The symbols represent a subset of simulation points in the weak capillary regime. The curve is obtained by fitting transition time divergence for various reduced volume, see figure \ref{fig_transitionTime_Ca}. Solid line and symbols represent the data for a vesicle released close to a wall $h/R_0=1.1$, while dashed line represent data for a vesicle in unbounded shear flow.}
\label{fig_beta_Ca}
\end{figure*}

Clearly, the dimensionless time needed to switch from the oblate state to the prolate state depends on the capillary number, as it increases with decreasing capillary number. Because characterizing the nature of the transition is not a straightforward task, we will later use this dependency to determine the type of bifurcation. We define the transition time $t_{tr}$ as the instant such that $\beta(t_{tr})=0$. This transition time and the critical capillary number depend on the reduced volume of the vesicle. In figure \ref{fig_transitionTime_Ca}-a, we represent the evolution of the transition time as a function of the capillary number. In the high $Ca$ regime, one could expect that the transition time decays to zero. However, because we always start our simulations with a symmetric shape whose axis are aligned with the wall, the vesicle has to first tilt towards the maximal stretching axis (located at $\pi/4$ in the xy-plane), before transiting from oblate to prolate. We note $t_{\infty}$ the time needed to reorient the shape in flow, and use it as a fitting parameter to adjust our curves. On the other hand, at low capillary number, this transition time diverges at a critical capillary number as $t_{tr} \sim \frac{1}{\sqrt{\Ca-\Ca_c}}$. This divergence is characteristic of a saddle-node transition. Note that the nature of the transition can be understood intuitively: for a given reduced volume $v < 0.75$, without external shear, two families of solutions exist, corresponding to local minima of the bending energy. Because bending energy is a continuous function of the deformation, these minima are connected by a local maximum, which is a stationary unstable point. Energy can be considered as a double-well curve of a state quantity, named $x$. The transition occurs when this unstable branch collides with the stable oblate branch. Indeed, increasing the capillary number $Ca$ decreases the oblate well up to vanish for $Ca = Ca_c(v)$. Close to this point, dynamics of oblate solution is governed by the dimensionless normal form of saddle-node transition:
\begin{equation}
\dot{x}=r\,+\,x^2
\end{equation}
where $r$ is proportional to $Ca - Ca_c(v)$. In the case $r < 0$, there are two fixed points, one stable and another unstable: $x = \pm \sqrt{-r}$. Otherwise, there is no solution: the oblate solution does not exist anymore. Scaling variables such that $x=\sqrt{r} X$ and $t = T / \sqrt{r}$ leads to: $\dot{X}=1\,+\,X^2$ which does not depend on $r$ and is straightforward to integrate between two quantity states. Thus, the necessary time scales as $1/\sqrt{r}$, a characteristic of saddle-node transition.

Note that the transition as function of reduced volume for a fixed non zero capillary number is also a saddle-node transition, as shown in the figure \ref{fig_transitionTime_Ca}-b : the evolution of the transition time is well-described by $t_{tr}\sim \frac{1}{\sqrt{v-v_c}}$.

To further characterize the transition, we also measure the amplitude of the shape parameter as a function of the capillary number, as shown in figure \ref{fig_beta_Ca}-a. In the range of studied reduced volume ([0.592, 0.75]), a prolate branch of solution exists for all capillary numbers, while the oblate branch of solution exists only for weak flows, and ceases to exist discontinuously at $\Ca=\Ca_c$. Rescaling $Ca$ as $Ca_c$ and $\beta$ as $\beta/\beta_{max}$ allows the collapse of the transition curves for three different reduced volumes. We have checked the sensitivity to perturbations in the subcritical case $Ca < Ca_c$ where both prolate and oblate are local minima of bending energy. The initial oblate shape is elongated in the direction of flow preserving volume, area and $\beta < 0$ (oblate shape). In the case $v=0.63$, we obtain a transition oblate to prolate for an initial deformation of $1\,\%$ and $Ca/Ca_c = 0.5/0.515 \approx 0.97$ and for an initial deformation of $4\,\%$ and $Ca/Ca_c = 0.375/0.515 \approx 0.73$: see two green asteriks on prolate branch (upper branch) in figure \ref{fig_beta_Ca}-a. The smaller the capillary number, the larger the initial perturbation.

This oblate to prolate transition is characteristic of shear flow and the value of critical capillary number is moderately affected by the presence of a wall: for a reduced volume of $v=0.59$, the critical capillary number in wall-bounded shear flow is $\Ca_c \approx 0.88$, while it is $\Ca_c^{FS} \approx 1.05$ in free-space. There is thus a roughly $15\%$ decrease in the critical capillary number due to the presence of a wall at $h=1.1 R_0$. The reduction in capillary number is due to the lubrication layer which holds the bottom part of the vesicle while the upper part is lifting: there is thus an additional stretching of the vesicle which triggers the transition sooner.


Finally, we also investigate how this transition is affected by the reduced volume.
The oblate shape is the global minimum \cite{Seifert_1991} of the bending energy  for reduced volume in the range $[0.592,0.651]$. For higher reduced volume, for which the prolate shape is the global minimum, discocyte solutions can still exist for small capillary numbers, indicating that the oblate familly is metastable even under weak flows. For a given reduced volume, the critical capillary number $Ca_c$ is determined by the fit presented in figure \ref{fig_transitionTime_Ca}. The evolution of $Ca_c$ as a function of reduced volume is presented in figure \ref{fig_beta_Ca}-b, together with a subset of the simulated points, indicating the zone where two solutions may be found, depending on the initial condition, and labelled as ``bistable''. The parameters where only prolate shapes were found (irrespective of the initial conditions) are labelled ``prolate''. As can be seen in figure \ref{fig_beta_Ca}-b, there is a good agreement between the zones determined by simulation points alone and the zones determined with the fit of transition time divergence. We also note that the range of possible capillary numbers compatible with oblate shapes decreases with increasing reduced volume: for $v=0,59$, the critical capillary number is around $0.9$ while for $v=0.725$ it is only $0.07$. For a reduced volume higher than $0.74$, we observe that a small amount of hydrodynamic stress is sufficient to trigger the oblate to prolate transition. This is in good agreement with the energetic stability analysis which shows that oblate shapes become a local minimum for $v < 0.75$ \cite{Jaric_1995}. For $v=0.648$, we find a critical capillary number of $Ca_c=0.44$. This capillary number is consistent with the range $[0.6;1.0]$ given by \cite{Spann_2014} for similar conditions ($v=0.65$,$\lambda=1$) because the definition of capillary number is based on a different length scale: \cite{Spann_2014} uses the radius of the sphere of equivalent area as the length scale, while we use the radius of the sphere having the same volume as the length scale. With the same definition of capillary number than \cite{Spann_2014}, we find a critical capillary number of $0.68$.

\section{Lift dynamics}

\begin{figure*}
\center
\includegraphics[scale=0.8,trim=0 335 300 0, clip=true]{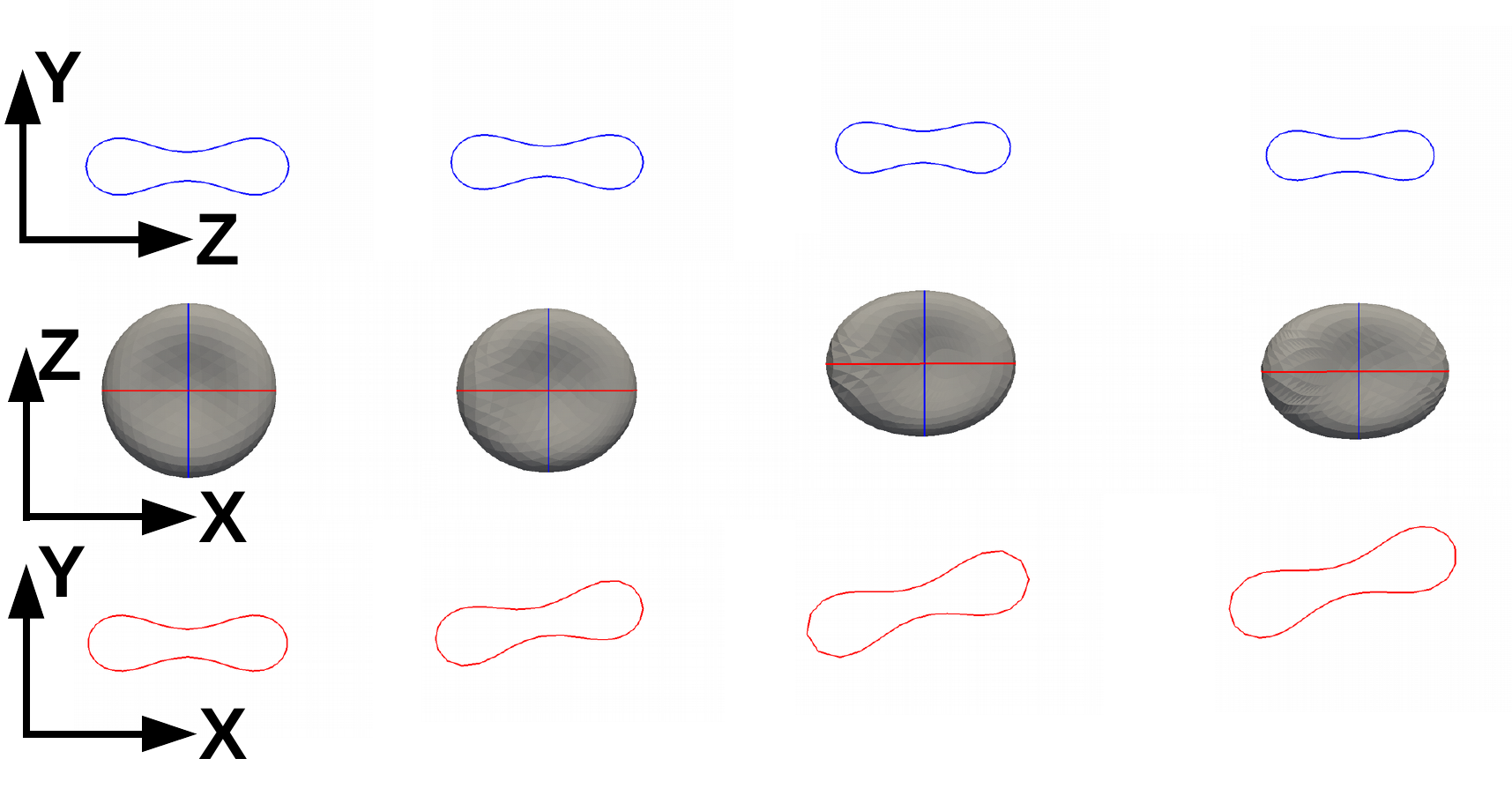}
\caption{Dynamics of lift of a biconcave vesicle ($v=0.635; Ca=0.375$): Snapshots in the $Z-Y$ plane (top row), $X-Z$ plane (middle row) and $X-Y$ plane (bottom row), for different dimensionless times (from left to right $\dot{\gamma}t = 0; 2; 5; 10$.}
 \label{fig_lift_biconcave}
\end{figure*}

\begin{figure*}
\begin{tabular}{cc}
\includegraphics[scale=0.7]{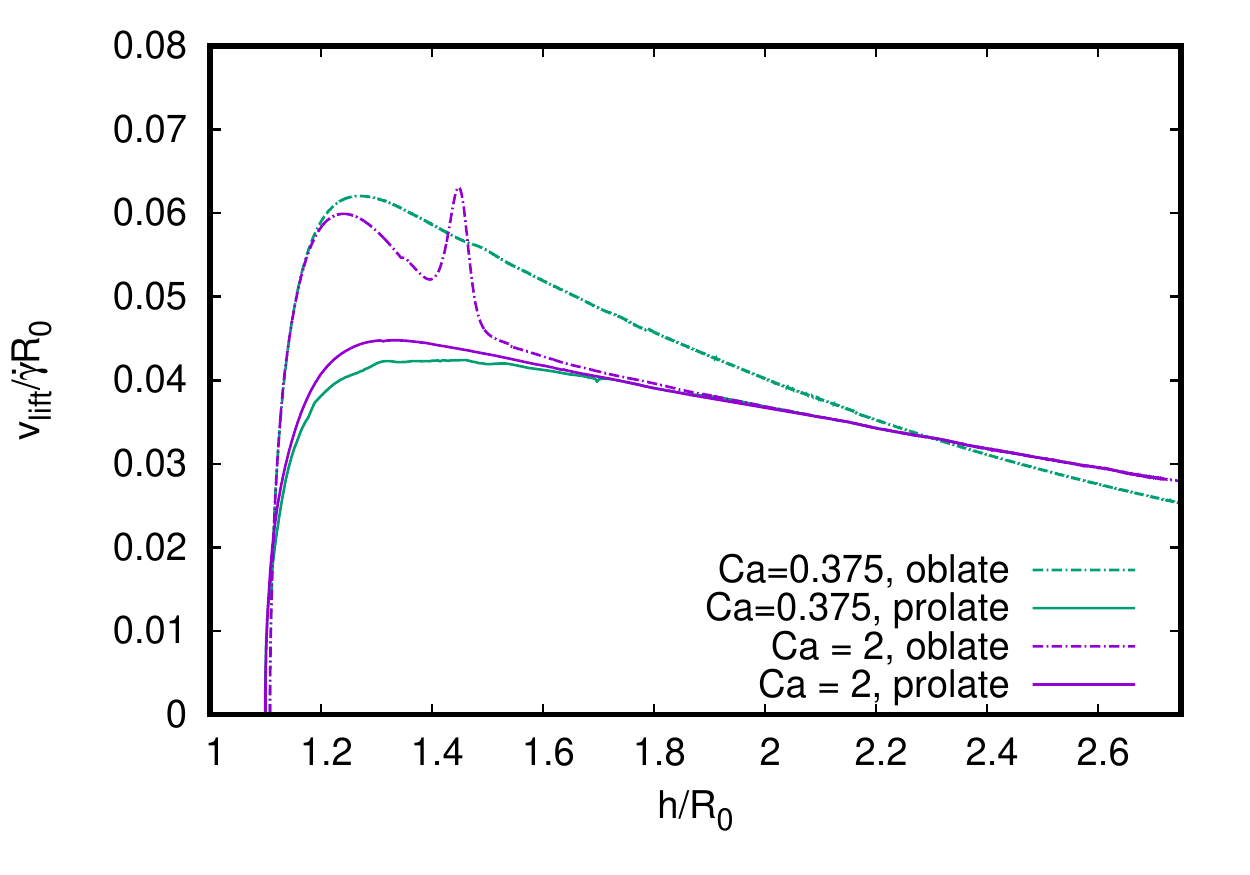}&
\includegraphics[scale=0.7]{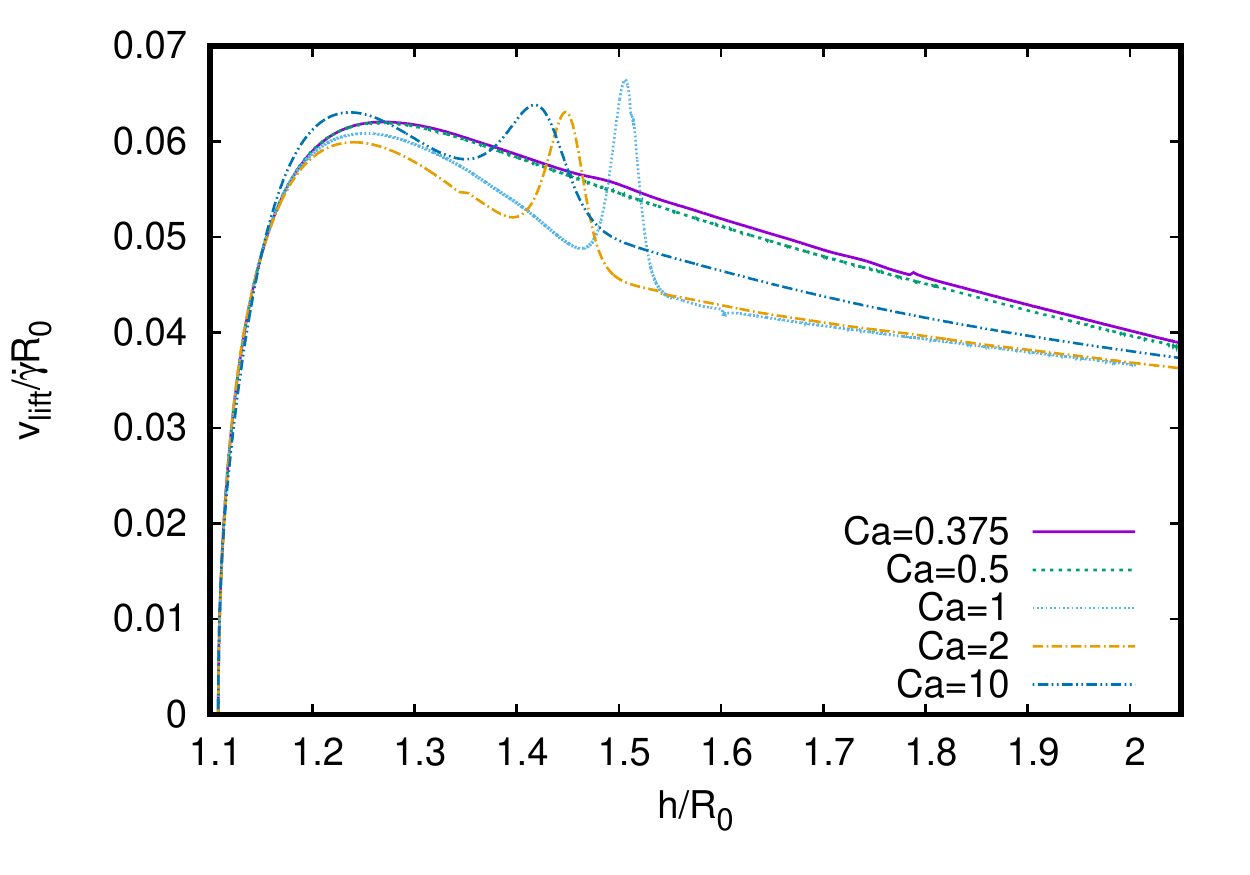}\\
(a) & (b)\\
\end{tabular}
\caption{(a) - Migration dynamics of a v=0.635 vesicle, as a function of initial shape (oblate - dashed lines, prolate - solid lines) and capillary number. Close to the wall, an oblate vesicle has a higher lift velocity, while a prolate vesicle has an higher lift velocity far from the wall. (b) - Dimensionless lift velocity as a function of dimensionless centroid height, for a v=0.635 vesicle, starting with an oblate shape. Critical capillary number for this reduced volume is $\approx 0.52$. The transition from oblate to prolate shape is visible in the lift velocity as ``bump'', whose relative importance is higher closer to the transition (see e.g. $Ca=1$ vs $Ca=10$ curves).}
\label{fig_migration_dynamics}
\label{fig_lift_function_height}
\end{figure*}
\begin{figure*}
\center
 \includegraphics[scale=0.65, trim=0 350 0 0, clip=true]{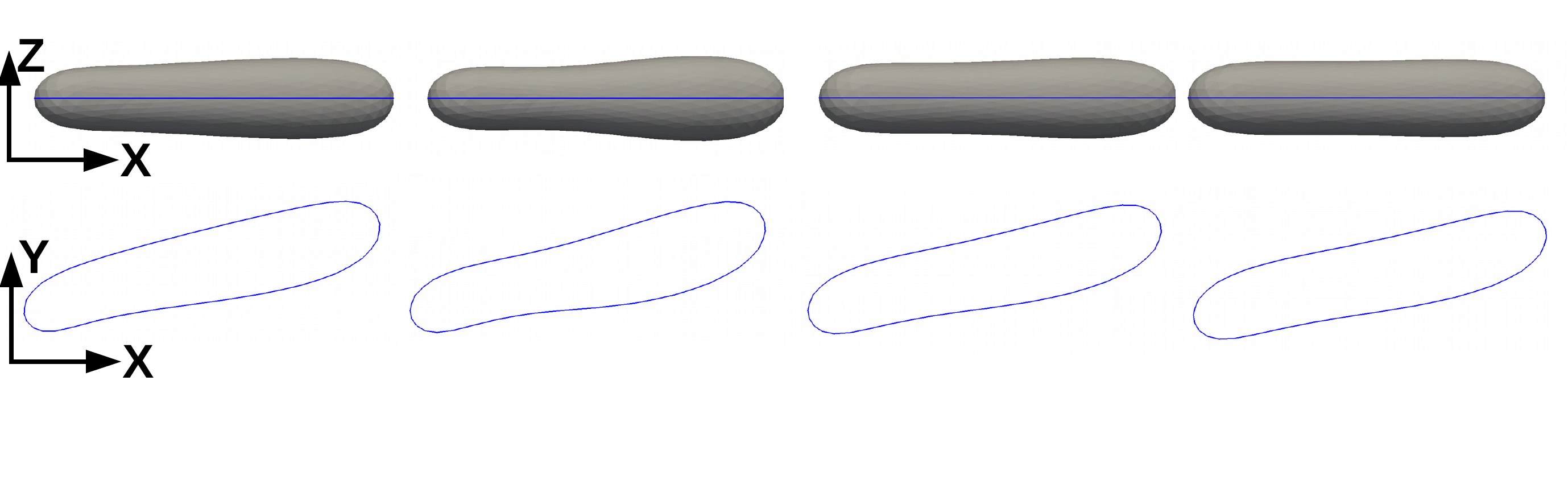}
 \caption{Asymmetric shape of a prolate vesicle ($Ca=10$). Snapshots in the $X-Z$ plane (top row) and $X-Y$ plane (bottom row), corresponding to dimensionless heights of $\frac{h}{R_0}=1.6;2.3;3.5;5$ }
 \label{fig_asym_shapes}
\end{figure*}

A typical simulation is shown in figure \ref{fig_lift_biconcave} : a discocyte shaped vesicle is placed in a weak shear flow ($Ca=0.375$). Because of the fore-aft symmetry of the initial shape, the initial lift velocity is zero. The shape of the vesicle adapts itself by stretching in the flow direction, while the membrane also develops a tank-treading motion. There is thus a pressure gradient in the lubrication film, the highest pressure being found at the front of the vesicle \cite{Cantat_1999,Seifert_1999}. This induces a torque which is balanced by the torque acting on the vesicle when its longest axis changes its orientation in flow, which leads to a reorientation of the vesicle with a positive tilt, as shown in figure \ref{fig_scheme}. In a second phase, the vesicle moves away from the wall, keeping roughly a constant deformation and tilt angle. This deformation measured by Taylor parameter $D_{xy}$ and orientation angle $\theta$ depends on the capillary number and the reduced volume. In the weak capillary regime (where oblate shapes are stable), increasing the capillary number increases the deformation $D_{xy}$ defined in figure \ref{fig_scheme} while the angle of inclination $\theta$ decreases. At larger capillary numbers, the evolution is reversed: $D_{xy}$ is a weakly decreasing function of the capillary number, while $\theta$ is a weakly increasing function of $Ca$.
Rescaling the lift velocity by $\dot{\gamma}$, all curves for weak flow collapse on a master curve $v_{lift}=v_{lift}(h)$ which depends only on the vesicle height $h$. Far enough from the wall, the asymptotic regime is $v_{lift}\sim h^{-2}$ with a prefactor depending on the particle stresslet as shown in \cite{Zhao_Pof2011}. In \cite{Zhao_Pof2011}, the authors show that, in this asymptotic regime, a prolate vesicle has always a higher lift velocity, because the amplitude of its particle stresslet is higher. However, close to the wall, the situation is different: while oblates still have a lower stresslet, they now have a (roughly $50 \%$) higher lift velocity than prolate ones, as shown in figure \ref{fig_migration_dynamics}-a. Note that, in the vicinity of the wall, lift velocity is dominated by lubrication effects. An oblate vesicle, having an higher area facing the wall than a prolate one, leads to a more efficient lubrication layer beneath the vesicle, which in turn leads to an higher lift velocity.

\begin{figure}
 \includegraphics[scale=0.14, trim=220 300 50 120, clip=true]{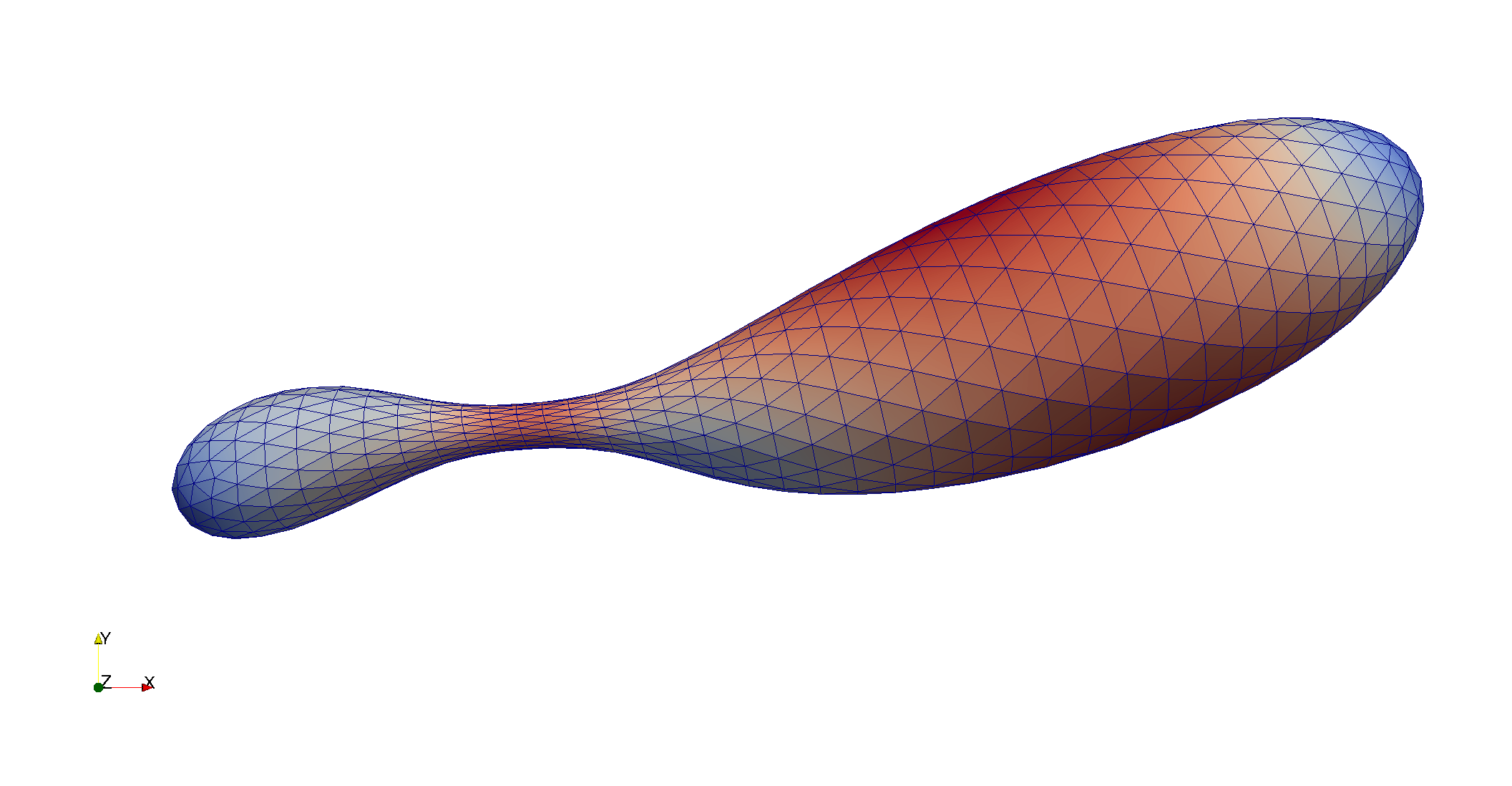}
 \caption{Transient tethered shape due to the presence of wall, obtained for a deflated vesicle without viscosity contrast ($\lambda=1$) under strong shear flow ($Ca=50$).}
 \label{fig_transientPB}
\end{figure}

Upon further increase of the capillary number, the initially oblate shape is stretched by the flow into a prolate shape, as discussed previously, and shown in figure \ref{fig_migration_dynamics}-b. This shape transition has also a signature in the lift velocity, which shows a local ``bump'' characteristic of the transition. The closer to the transition, the higher the local ``bump'' is, as shown in figure \ref{fig_lift_function_height}. After the transition, the lift velocity is  the same, irrespective of the initial shape, as shown in figure \ref{fig_migration_dynamics}-a for $Ca=2$.




Finally, for high ($Ca \ge 10$) capillary numbers, the migration from the wall is accompanied by the appearance of transient asymmetric shapes, as shown in figure \ref{fig_asym_shapes}. Far enough from the wall, the asymmetry vanishes as the shape converges towards the free-space one. Increasing the capillary number leads to a more pronounced asymmetry, which might even lead to the formation of a transient tethered shape, as preliminary simulations at higher capillary numbers ($\approx 50$) indicate: figure \ref{fig_transientPB}. These shapes are reminiscent of the PB state obtained by \cite{Farutin_2012} for a deflated vesicle ($v\approx 0.6$) in shear flow with a viscosity contrast ($\lambda < 1.0$). Note that the appearance of this transient tethering is not related to the oblate-prolate transition, as it appears independently of the initial shape when starting either with an oblate or a prolate. It is due to the asymmetry in the hydrodynamic stresses applied to the vesicle : the presence of the wall leads to an additional stress in the lubrication layer. This tends to 'pin' the vesicle near the wall, while the upper part is lifted and stretched by the flow.


%
%

\section{Conclusion}

We studied how hydrodynamic stresses can trigger the transition between an oblate shape and a prolate shape. We provide an accurate diagram showing in which region of the ($v,Ca$) plane bistable solutions may be found. For quasi-spherical vesicle, no bistable region was found. For more deflated vesicles, a bistable region was found for weak flow, with an upper bound increasing when decreasing the reduced volume. Note that for deflated vesicles ($v\sim 0.6$) viscosity contrast \cite{Spann_2014} or membrane viscosity \cite{Noguchi_2004,Noguchi_2005} were both found to increase the critical capillary number, thus stabilizing the oblate shape. 
The presence of the wall leads to a slightly decreased critical capillary number, due to the additional stress. We then studied lift dynamics of oblate vesicle in the weak flow regime and found that, contrary to the far-field prediction, oblate vesicle near the wall have an higher lift velocity when compared to prolate vesicle with the same reduced volume. Finally, we showed that, for high capillary numbers, wall-induced hydrodynamic stresses may also lead to asymmetric shapes, independently of the initial shape.

\section{Acknowledgements}
This work has benefited from financial support from the ANR 2DVISC (grant no. ANR-18-CE06-0008), from Labex MEC (grant no. ANR-11-LABX-0092), from A*MIDEX (grant no. ANR-11-IDEX-0001-02) and from CNES. This work was granted access to the HPC resources of Aix-Marseille Universit\'e financed by the project Equip@Meso (ANR-10-EQPX-29-01) of the program ``Investissements d'Avenir'' supervised by the Agence Nationale de la Recherche. M.D. was supported financially by a fellowship cofunded by the region ``Provence-Alpes-C\^ote-d'Azur'' and the CNRS - INSIS. LRP is part of the LabEx Tec 21 (Investissements d'Avenir -grant agreement ANR-11-LABX-0030) and of the PolyNat Carnot Institute(Investissements d'Avenir -grant agreement ANR-11-CARN-007-01).

\section{Authors contributions}
MD has performed the numerical results. GB and ML were involved in the preparation of the manuscript.
All the authors have read and approved the final manuscript.
%

\begin{thebibliography}{}
\bibitem{BarthesBiesel_2016}
D. Barth\`{e}s-Biesel, Ann. Rev. Fluid Mech. \textbf{48}, 25 (2016)
\bibitem{Dimova_2002}
R. Dimova, U. Seifert, B. Pouligny, S. F\"{o}rster, H. G. D\"{o}bereiner, Eur. Phys. J. E \textbf{7}, 241 (2002)
\bibitem{Discher_2006}
D. E. Disher, F. Ahmed, Ann. Rev. Biomed. Eng. \textbf{8}, 323 (2006) 
\bibitem{Meng_2009}
F. Meng, Z. Zhong, J. Feijen, Biomacromolecules \textbf{10}, 323 (2006)
\bibitem{Kantsler_2005}
V. Kantsler, V. Steinberg, Phys. Rev. Lett. \textbf{95}, 258101 (2005)
\bibitem{DeLoubens_2016}
C. De Loubens, J. Deschamps, F. Edwards-L\'evy, M. Leonetti, J. Fluid Mech. \textbf{789}, 750 (2016)
\bibitem{Deschamps_2009}
J. Deschamps, V. Kantsler, V. Steinberg, Phys. Rev. Lett. \textbf{102}, 118105 (2009)
\bibitem{Goldsmith_1962}
H. L. Goldsmith, S. G. Mason, JCIS \textbf{17}, 448 (1962)
\bibitem{Smart_1991}
J. R. Smart, D. T. Leighton Jr, Phys. Fluids A \textbf{3}, 21 (1991)
\bibitem{Olla_1997}
P. Olla, J. Phys. II, \textbf{7}, 1533 (1997)
\bibitem{Vlahovska_2007}
P. M. Vlahovska, R. S. Garcia, Phys. Rev. E \textbf{75}, 016313 (2007)
\bibitem{Farutin_2013}
A. farutin, C. Misbah, Phys. Rev. Lett. \textbf{110}, 108104 (2013)
\bibitem{Zhao_Pof2011}
H. Zhao, A. P. Spann, E. S. G. Shaqfeh, Phys. Fluids \textbf{23}, 121901 (2011)
\bibitem{Callens_2008}
N. Callens, C. Minetti, G. Coupier, M. -A. Mader, F. Dubois, C. Misbah, T. Podgorski, EPL \textbf{83}, 24002 (2008)
\bibitem{Cantat_1999}
I. Cantat, C. Misbah, Phys. Rev. Lett. \textbf{83}, 880 (1999)
\bibitem{Seifert_1999}
U. Seifert, Phys. Rev. Lett. \textbf{83}, 876 (1999)
\bibitem{Sukumaran_2001}
S. Sukumaran, U. Seifert, Phys. Rev. E \textbf{64}, 011916 (2001)
\bibitem{Lorz_2000}
B. Lorz, R. Simson, J. Nardi, E. Sackmann, EPL \textbf{51}, 468 (2000)
\bibitem{Abkarian_2002}
M. Abkarian, C. Lartigue, A. Viallat, Phys. Rev. Lett. \textbf{88}, 068103 (2002)
\bibitem{Abkarian_2005}
M. Abkarian, A. Viallat, Biophys. J. \textbf{89}, 1055 (2005)
\bibitem{Messlinger_2009}
S. Me{\ss}linger, B. Schmidt, H. Noguchi, G. Gompper, Phys. Rev. E \textbf{80}, 011901 (2009)
\bibitem{Kraus_1995}
M. Kraus, U. Seifert, R. Lipowsky, EPL \textbf{32}, 431 (1995)
\bibitem{Seifert_1991}
U. Seifert, K. Berndl, R. Lipowsky, Phys. Rev. A \textbf{44}, 1182 (1991) 
\bibitem{Seifert_1997}
U. Seifert, Adv. Phys. \textbf{46}, 13 (1997)
\bibitem{Spann_2014}
A. P. Spann, H. Zhao, E. S. G. Shaqfeh, Phys. Fluids \textbf{26}, 031902 (2014)
\bibitem{Noguchi_2004}
H. Noguchi, G. Gompper, Phys. Rev. Lett. \textbf{93}, 258102 (2004) 
\bibitem{Noguchi_2005}
H. Noguchi, G. Gompper, Phys. Rev. E \textbf{72}, 011901 (2005
\bibitem{Blake_1971}
J. Blake, Mathematical Proceedings of the Cambridge Philosophical Society  \textbf{70}, 303 (1971)
\bibitem{Boedec2011}
G. Boedec, M. Leonetti, M. Jaeger, J. Comp. Phys. \textbf{}, (2011)
\bibitem{Zhao2011}
H. Zhao, E. S. G. Shaqfeh, J. Fluid Mech. \textbf{674}, 578 (2011)
\bibitem{Biben2011}
T. Biben, A. Farutin, C. Misbah, Phys. Rev. E \textbf{83}, 031921 (2011)
\bibitem{Loubens2015}
C. De Loubens, J. Deschamps, G. Boedec, M. Leonetti, J. Fluid Mech. \textbf{767}, R3 (2015)
\bibitem{Gounley2016}
J. Gounley, G. Boedec, M. Jaeger, M. Leonetti, J. Fluid Mech. \textbf{791}, 464 (2016)
\bibitem{Boedec_2017}
G. Boedec, M. Leonetti, M. Jaeger, J. Comp. Phys. \textbf{342}, 117 (2017)
\bibitem{Jaric_1995}
M. Jari\'c, U. Seifert, W. Wintz, M. Wortis, Phys. Rev. E \textbf{52}, 6623 (1995)
\bibitem{Farutin_2012}
A. Farutin, C. Misbah, Phys. Rev. Lett. \textbf{109}, 248106 (2012)



\end{thebibliography}
%

\end{document}